\documentclass[aps,showpacs,amsmath,amssymb,prl, twocolumn]{revtex4}
\usepackage{graphicx}
\usepackage{dcolumn}
\usepackage{bm}
\usepackage{hyperref}
\usepackage{epsfig}
\usepackage{wasysym}
\begin{document}
\title{On a Second Critical Point in the First Order  Metal - Insulator Transitions}
\author{Ivan Z. Kostadinov, Bruce R. Patton}
\email{kostadinov.1@osu.edu}
\affiliation{Department of Physics, The Ohio State University, Columbus OH 43210}
\date{\today}
\begin{abstract}
    For the first order Metal Insulator Transitions we show that
together with the d.c conductance zero there is a second critical
point, where the dielectric constant becomes zero and further
turns negative. At this point the metallic reflectivity sharply
increases. The two points can be separated by a Phase Separation
State in a 3D disordered system, but may tend to merge in 2D. For
illustration we evaluate the dielectric function in a simple
effective medium approximation and show  that at the second point
it turns negative. We reproduce the experimental data on a typical
Mott insulator like MnO, demonstrating the presence of the two
points clearly. We discuss other experiments for studies of the
phase separation state and a similar phase separation in
superconductors with insulating inclusions.
\end{abstract}
\maketitle
    The charge transport and Metal Insulator Transitions (MIT) in disordered
systems have been discussed in many papers
\cite{PWA,gangof4,Finkelstein}, reviews\cite{Lee,review-mit} and
books\cite{Mott,Altshuler, Stauffer, Bergman}. The initial
discussion resulted in agreement that  at $T=0$ the transition is
completely described by the dc conduction as a function of the
Fermi level at $T=0$. This is true for second order MIT, where
simultaneously with the vanishing of the conductance vanishes also
the real part of the dielectric function. It is negative in the
metallic state, positive in the dielectric one and there is no
phase separation in this case. Here we show that in a first order
MIT in composites the dc conduction is not enough and the dc
dielectric constant has to be considered in order to make the
description complete. We will show that in a first order MIT in
the composites, in disordered thermodynamically metastable 3-D
systems, there can be an intermediate two phase region separating
the metallic state from the dielectric one, both at $T=0$ and at
finite temperatures. We will consider such a metal - dielectric
composite with a metallic volume fraction $f$, but our results are
relevant for other first order MIT in homogeneous systems, due to
the nucleation of metallic inclusions in the dielectric matrix
under pressure for example. In a composite, grains of the both
components are always present for any metallic volume fraction $f
\neq 0$ and $f \neq 1$ and the resistance drop occurs at a volume
fraction $f_{c}$ different from the dielectric function sign
change $f_{d}$ and in different samples made with different
metallic volume fractions. Under pressure due to different
compressibility a composite irreversibly can undergo both
transitions. To illustrate this we show a non-composite example of
the phase separation in the Mott insulator MnO \cite{kondo-x-ray},
\cite{patterson}, \cite{MnO}. It demonstrates both the resistance
drop near 90 Gpa and subsequent reflectance sharp increase, which
becomes pressure independent at 127GPa\cite{reflectance-MnO}. The
data from\cite{MnO} are given in Fig.{\ref{fig:exp}}and a detailed
discussion of MIT in MnO is given in \cite{kunes-nature}.
\section{Specific Phase Separation State}
     A metallic 3D sample is screening any static electric
fields as well as the electron-electron interaction. Let $f_{c}$
be the metallic volume fraction threshold for the first non zero
conductance to appear. From the percolation theory is known that
often the values of  $f_{c}$ for 3-D systems are smaller than 0.5
and are different for various models \cite{Stauffer} We will prove
now for 3-D composites that in a first order MIT in the range of
values $f>f_{c}$ the sample can be conducting but not screening
the external static electric fields. Just above $f_{c}$ the sample
is in a larger part made of dielectric grains, which form an
infinite cluster of a larger volume than the volume of the
metallic infinite cluster, provided that $f_{c}< 0.5$. \emph{The
two coexisting infinite clusters represent a specific type of
phase separation state}. Due to the topological structure of the
infinite clusters in a disordered system such phase separation is
clearly different from a layered one for example. When placed in
the external static electric field of a capacitor, such a sample,
being in a large part dielectric, becomes polarized with non-zero
volume polarization. This volume polarization requires energy for
creating the positive energy density in the polarized infinite
dielectric cluster, which is equal to $\frac{\varepsilon{\rm
E}^{2}}{8\pi}$, where $\varepsilon$ is the real part of the static
(zero frequency) dielectric constant of the sample. The real part
of $\varepsilon$ is certainly positive due to the work done for
polarizing the sample and not negative as it is in a metal. Thus,
we have demonstrated that the phase with both, conductance and
volume polarization is not a metal, nor a dielectric and
represents a specific phase separation state. The data presented
in Fig.{\ref{fig:exp}} show a separation state as a result of a
first order MIT in a different system - MnO, which is not a
composite and the metallic phase appears gradually due to
nucleation.The resistance drop is near 94 GPa (phases dB1+B8 in
Fig.{\ref{fig:exp}}, B1 for rocksalt, dB1 for a rhombohedral
distortion of B1, and B8 for NiAs)). The further increase of the
metallic volume fraction $f$ breaks the last links between the
large dielectric clusters and all dielectric inclusions become
eventually disconnected near 127 GPa. Near this point the sample
becomes a metallic bulk with isolated dielectric inclusions. The
bulk metal has negative dielectric constant and at this point (the
disappearance of the infinite dielectric cluster) the static
dielectric constant turns from positive to negative. The
reflectance is pressure independent above 127 GPa in the bulk
metal region.  Due to the tunnelling of the electrons(holes) and
the hybridization of the wave functions of the large dielectric
clusters the real critical point is at values different from the
geometric dielectric percolation threshold. To illustrate this
second critical point we applied here a simple version of the
Effective Medium Approximation (EMA) to evaluate the dielectric
function. Let us mention that EMA is limited in precision and is
one among many similar ones differing in the
details\cite{Bergman,EMA}.
\section{Experimental characterization}
    Experimentally one can expect in a first order MIT the metallic
reflectance to decrease  sharply below this second critical point
$f_{d}$, which is different from the sharp resistance drop point
$f_{c}$. This is also the case with the observed metallic luster
in the successive measurements of the same material
MnO\cite{kondo-x-ray,patterson,MnO}. Therefore below the critical
point $f_{d}$ the reflectivity decreases sharply due to the
transmission of the infinite dielectric cluster and specific
plasmons passing through the connected system of tunnels in the
metal. The d.c. polarization $P$ should appear below the point
$f_{d}$ in static electric field in a capacitor and experiments on
its critical behavior are of considerable interest as well as
experiments on the behavior of the ac currents dissipation in the
region, where infinite clusters emerge or disappear. The small
dielectric inclusions in the metal will reduce eddy currents
dissipation at small frequencies, like in external time dependent
magnetic field. When the metal concentration decreases and becomes
$ f\leq f_{d}$, this eddy currents dissipation would have a
feature due to the change in the topology of the dielectric
inclusions and the appearance of the infinite dielectric cluster,
which also marks the bulk metallic volume reduction. Thus the eddy
currents signal of a coil will show a variation near the second
critical point. Another experiment can be carried out in the phase
separation state, where one may encounter different sound
velocities related to the two infinite clusters with different
type of bonding\cite{Hg}. In general, we have shown that the
dielectric function  have a second critical point at $f_{d}$,
where the real part of the static $\epsilon$ becomes negative
entering the metallic state region. When this critical point does
not coincide with the vanishing  dc conduction there will be an
intermediate two phase region. In a quasi 2D system made out of
parallel infinite metallic cylinders such phase separation does
not exist at all. In a first order MIT the phase separation region
was observed in various Mott insulators like in the MnO room
temperature MIT transition shown here.
\begin{figure}[t]
\includegraphics[width=\columnwidth]{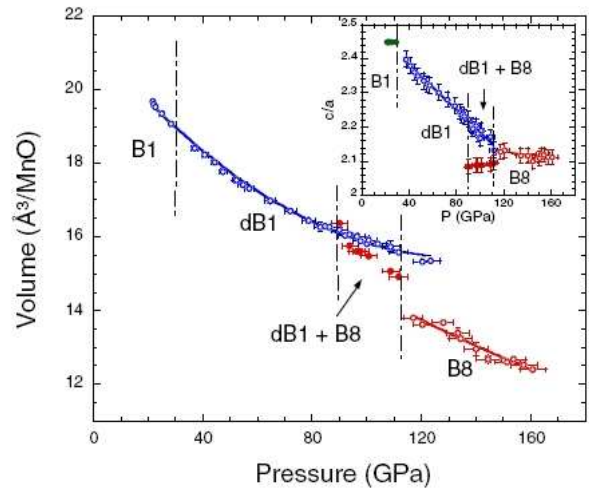}
\caption{\emph{from ref.} \cite{MnO}The specific volume and the
c/a ratio (inset) of MnO phases(B1-rocksalt,dB1-a rhombohedral
distortion of B1, and B8 for NiAs) as a function of pressure. Note
the discontinuous changes of the specific volume and the c/a ratio
at 110 GPa, indicating that MnO undergoes an isostructural phase
transition with $6.6 \%$ volume collapse. This transition
coincides within experimental uncertainties with the moment loss
and the insulator-metal transition in \cite{patterson}.}
\label{fig:exp}
\end{figure}
    The experimental data of reference\cite{MnO} reproduced in the
Fig.{\ref{fig:exp}} present a volume range in the pressure -
volume diagram separating the metallic and the insulating phases
of MnO and showing $6.6$ \% volume difference. It corresponds to
the difference in the metallic volume fraction between the two
critical points mentioned above for the case of the composites.
 \section{Two types of extended states}
    What describes the phase separation state in the range of
concentrations$f_{c}\leq f \leq f_{d}$? The metallic phase is well
defined in the metallic side $f_{d}\leq f \leq 1$ and the
insulating phase is as well defined (see further) in the small
metallic fraction region $0\leq f \leq f_{c}$. In the intermediate
range $f_{c}\leq f \leq f_{d}$ one have coexistence of the two
phases in the form of two infinite clusters.
    Remarkably in this region there is also coexistence of two different
types of extended states-metallic ones threading the infinite
metallic cluster and dielectric ones based on the infinite
dielectric cluster. In the dielectric type of extended states
between the top and bottom  mobility edges, an electron (hole),
when injected in the dielectric cluster can cross the sample as
well as it can in the metallic one. For $f \geq f_{d}$ in the
metallic phase only metallic type of extended states exists and
the sample conductivity is metallic. For $f \leq f_{c}$ in the
dielectric phase only dielectric type of extended states is
present and due to the gap, the sample is an insulator at $T=0$.
In the vicinity of $f_{c}$ at $T=0$  metallic component states
 go from localized to delocalized following  scaling laws
as described by the Localization
Theory{\cite{PWA,gangof4,review-mit}.
     In the Table.\ref{3states-table}, the static order parameters
values are displayed for all states. The symbol $\mathbf{P}$ is
the isotropic d.c. polarization value.
\begin{table}
\caption{\textbf{The order parameters of metal,insulator and phase
separation states}}
\begin{ruledtabular}
\begin{tabular}{lll}

\textbf{Insulator}
& \textbf{Phase separation state} & \textbf{Metal}\\
$\mathbf{\sigma_{d.c.} = 0 }$ & $\mathbf{\sigma_{d.c.} \neq 0}$
& $\mathbf{\sigma_{d.c.}\neq 0}$\\
$\mathbf{P\neq 0 }$ & $\mathbf{P\neq 0}$ & $\mathbf{P =0}$\\

\\
\end{tabular}
\end{ruledtabular}
\label{3states-table}
\end{table}
\section{The Second Critical Point}
    In this section we show how the second critical point
appears. The two infinite clusters  and the electronic states
based on metallic or dielectric grains  and clusters of grains are
determined quantum mechanically and not just geometrically. The
description also involves the polarization properties of both, the
metal and the insulator -$\Pi(\omega,f)$. Eventually the exact
form of the metal fraction $f$ dependence of the conductance and
the dielectric function and the values of the critical points can
vary depending on the models for the calculation.  We will use
Drude type form of $\Pi(\omega,f)$, which is appropriate for
illustration purposes. The EMA describes both critical points for
the conductance and for the dielectric function and this is enough
to demonstrate how the two critical points appear. A simple
version of (EMA) (e.g. not accounting for various shapes or
coatings of the grains)  for the conductivity$\sigma_{eff}$ of the
composite has the form \cite{Bergman} equation\label{Eq:effmed}:
\begin{equation}
\Sigma{f_{i}\frac{\sigma_{eff}-\sigma_{i}}{2\sigma_{eff}+\sigma_{i}}}
= 0
\end{equation}
When the first component  $i=1$ is  a metal and the second
component is an insulator this equation gives us the first
critical point $f_{c} = 1/3$, which is usually assumed to be the
point where the metal insulator transition happens. Next we find
the static dielectric constant \cite{Stroud} for the same system
using the same equation as above, but with $\varepsilon_{eff}$
instead of  $\sigma_{eff}$ and $\varepsilon_{1}$ for the metal,
the dielectric being marked $i=2$. We neglect here all spectral
behavior of the dielectric (and the metal) by assuming a
constant$\varepsilon_{diel}$ . For an \emph{ideal metal the static
dielectric constant is infinite and negative}. The above equation
gives us $\varepsilon_{eff}$ of an ideal metal in a dielectric in
the form:
\begin{equation}\label{Eq:pole}
   \varepsilon_{eff} = \frac{\epsilon_{diel}}{1-3f}
\end{equation}
Thus, \emph{the ideal metal in a composite with any insulator has
a static dielectric constant, which has a pole at the first
critical point $f_{c} = 1/3$.}  For a real metal in a composite
with a dielectric the pole is shifted up in the complex frequency
plane and turns into a maximum and eventually becomes negative. We
take the real metal dielectric function $\epsilon_{m}$ in the
Drude form:
 \begin{equation}
    \epsilon_{m}(\omega,f) = 1 - f\frac{\omega^{2}_{p}\tau(1 + i\omega\tau)
 }{i\omega(1+\omega^{2}\tau^{2})}
\end{equation}
    Here the plasma frequency $\omega_{p}$ is defined by the
electronic concentration of the metal, which is proportional to
the metal content $f$ . The relaxation rate $\tau^{-1}$ is
typically one order of magnitude less than the plasma frequency.
It may depend, however, strongly on the temperature or the
pressure. Using the EMA equation above for the effective
dielectric function of the composite we find
\begin{equation}
\epsilon_{eff}(\omega,f) = \frac{1}{4}[\alpha \pm
\sqrt{\alpha^{2}+ 8\epsilon_{d}\epsilon_{m}(\omega,f)}],
\end{equation}
Here the function $\alpha$ has the form:
\begin{equation}
    \alpha = (2-3f)\epsilon_{d} + (3f-1)\epsilon_{m}(\omega,f).
\end{equation}
    This formula is illustrated in Fig.{\ref{fig:criticalline}}, which
shows the real part of the dielectric function and the critical
line, where it becomes negative is clearly visible.
\begin{figure}[t]
\includegraphics[width=\columnwidth]{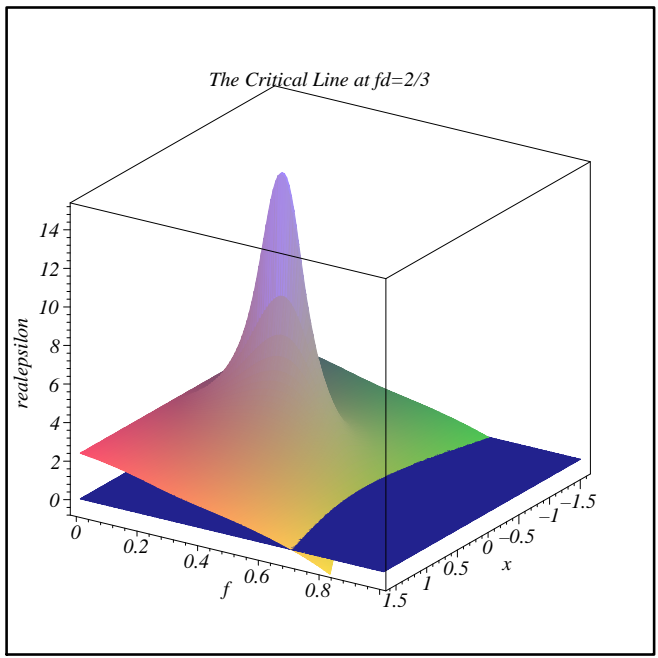}
\caption{The real part of the dielectric function as a function of
the metal volume fraction $f$ and the dimensionless frequency $ x
= \omega\tau $. The critical line at $ f_{d}=\frac{2}{3}$, is the
cross section of the blue colored plane defined as $ {\rm Re}
\varepsilon=0$. The dimensionless plasma frequency here is
$\omega_{p}\tau=13.7$. {\emph{The metallic phase is in the region
where the real part of the static dielectric function becomes
negative - below the horizontal blue plane}.}}
\label{fig:criticalline}
\end{figure}
    The maximum in the region around {$ f=\frac{1}{3}$} clearly is
related to the pole (see Eq.{\ref{Eq:pole}}) in the static
effective dielectric constant of a composite made out of an ideal
metal and a dielectric. For a real metal it turns into a finite
maximum as expected.
\section{The Superconductor Dielectric Composites}
    The treatment we applied for the intermediate state region is not
limited to the metal - insulator composites and MIT. It is easily
seen that the non-magnetic dielectric and superconductor
composites show similar behavior. Starting with a small volume
fraction of the superconductor f we see that the first
superconductive paths appear at the concentration $f_{c}$. The
magnetic susceptibility will sense the presence of the
superconductor component, but the complete screening of the
external magnetic field will occur only at $f\geq f_{d}$, when
there is no longer an infinite dielectric cluster present. In the
intermediate range of concentrations the obtained phase is a
superconductor and at the same time has a nonzero static
polarization as a dielectric in the range of temperatures $0\leq
T\leq T_{c}$. This picture is summarized in Table.{\ref{sc-diel}},
where we assume a diamagnetic dielectric. The interphase energy at
the two critical points is due to the expulsion of the electric
field penetrating the dielectric but in addition to it now there
is present the magnetic field energy density due to the magnetic
field screening, which may be partial near {$f_{c}$} and complete
at {$f_{d}$}.
\begin{table}
\caption{\textbf{The Dielectric Superconductor Composites}}
\begin{ruledtabular}
\begin{tabular}{lll}
\textbf{Dielectric} & \textbf{Phase Separation State} &\textbf{Superconductor}\\
$\mathbf{\sigma_{d.c.} = 0 }$ & $\mathbf{\sigma_{d.c.}=\infty }$
& $\mathbf{\sigma_{d.c.}=\infty}$\\
$\mathbf{P\neq 0 }$ & $\mathbf{P\neq 0}$ & $\mathbf{P =0}$\\
$\mathbf{\chi\leq 0 }$ & $\mathbf{\chi\leq 0 }$
& $\mathbf{4\pi\chi + 1 =0}$\\

\\
\end{tabular}
\end{ruledtabular}
\label{sc-diel}
\end{table}
The penetration depth $\lambda^{-2}$  is proportional to the
superfluid density $n_{s}$ and when it is zero the depth is
infinite. At low levels of doping the insulating phase of the
cuprates plays the role of the dielectric component. At $T=0$ K in
the still superconductive samples the magnetic field penetration
depth increases infinitely in the region of small doping
concentrations p less then $p=0.1$, the critical concentration
$p_{c}$ being close to it. Eventually the samples are
superconductive with low magnetic field penetration depth equal to
infinity. This absence of complete screening of the magnetic field
is marked in Table.{{\ref{sc-diel}}} as negative susceptibility,
but not vanishing $\mu = 1+ 4\pi\chi $.
    \textbf{Concluding}, we
have shown that in the first order MIT there can be a second
critical point, where $\varepsilon_{eff}=0$. Between the two
critical points there is a phase separation state, where two types
of extended states are present. In the same way in the
Superconductor Dielectric composites the dielectric phase and the
superconductive one can be also separated by an intermediate
phase. The second critical point where the infinite dielectric
cluster appears in the metallic phase is observable  by means of
static polarization studies, by magnetic susceptibility and by
eddy currents loss experiments, not to mention different sound
velocities  and the features in the reflection and/or transmission
of electromagnetic waves.\ \\
    We would like to thank Prof. V.L. Pokrovsky for stimulating
discussions and mentioning the recent work of the late A. M.
Dykhne \cite{Dykhne}. We also thank Dr A. McMahan and the authors
of \cite{MnO} for kindly giving us permission to reproduce their
results in Fig.\ref{fig:exp}.

\end{document}